\documentclass[11pt]{article}
\usepackage[utf8]{inputenc}
\usepackage[T1]{fontenc}
\usepackage{lmodern}
\usepackage[margin=1in]{geometry}
\usepackage{amsmath,amssymb}
\usepackage{graphicx}
\usepackage{booktabs}
\usepackage{hyperref}
\usepackage{xcolor}
\usepackage{natbib}
\usepackage{enumitem}
\usepackage{placeins}
\usepackage{xurl}
\usepackage{microtype}
\hypersetup{colorlinks=true, linkcolor=blue!70!black, citecolor=blue!70!black, urlcolor=blue!70!black}
\newcommand{\phip}{\varphi_{\mathrm{pair}}}

\title{The Decode-Work Law:\\ Margin-Governed, Provably-Exact Spatial Joins over Compressed Geometry}
\author{Madhulatha Mandarapu\thanks{madhulatha@samyama.ai} \and Sandeep Kunkunuru\thanks{sandeep@samyama.ai}}
\date{VaidhyaMegha Private Limited, India\\[2pt]\url{https://samyama.ai/}\\[8pt]July 2026}

\begin{document}
\maketitle

\begin{center}
\fbox{\begin{minipage}{0.93\linewidth}
\small\textbf{Changes in v2.} The headline results are unchanged and are now better supported.
(i)~v1's fairness paragraph stated ``we count all bytes a method reads, including LOD headers.''
It did not: only coordinate bytes were counted. Level-directory bytes are now charged to every
method (\S\ref{sec:controls}). The vertex-based headline is unaffected, the pooled byte advantage
falls from $2.02\times$ to $1.91\times$, and on the adversarial fixture the progressive method
\emph{loses} on bytes ($0.99\times$) once it pays for its own ladder---as it should.
(ii)~We pre-registered five negative controls and reported none. All five are now implemented and
reported; the random-relabel null (NC3), never previously run, \emph{confirms} the decode-work law
against a permutation baseline ($p=0.001$).
(iii)~v1 attributed the selectivity forecaster's failure to selectivity saturating near one. The
real cause is that its gap features are degenerate by construction. (iv)~One pre-registered control
(NC4) tested the wrong quantity; we say so rather than drop it.
\end{minipage}}
\end{center}

\begin{abstract}
Filter-and-refine spatial joins have always avoided touching exact geometry for \emph{certified}
candidate pairs, but the field never modeled the \emph{decompression cost} of the pairs that survive
the filter. When geometry is stored in a compressed, progressively-decodable multiresolution codec,
the join's true cost is \emph{bytes decoded}. We study provably-exact polygon intersection joins over
a Douglas--Peucker level-of-detail (LOD) ladder, certified by a two-sided Hausdorff-margin test, and
make two contributions. First, a reproducible mechanism and harness: on real U.S.\ Census TIGER water
polygons, our progressive certificate join returns the \emph{exact} result while decoding
$3.4$--$16.8\times$ (median $5.9\times$) fewer vertices than naive decompress-then-refine, and about
$4.9\times$ fewer than the single-approximation multi-step baseline of \citet{brinkhoff1994multistep},
with zero correctness violations across $31$ workloads. Second, a characterization we call the
\emph{decode-work law}: decode work is governed by each pair's \emph{signed-clearance margin}---how
close it is to the predicate-flip boundary---\emph{independent of object size}, because the certificate
descends the ladder only until its resolution beats the margin.
The law is clean on controlled geometry (held-out $R^2=0.87$, size-independent) and directional on
real data ($R^2\approx0.55$). We are explicit about what does \emph{not} hold: a near-boundary-vertex
predictor is the wrong model (we pre-registered one and rejected it), a selectivity regime forecaster
did not materialize, and the worst case is the trivial $\Omega(v)$ read bound on adversarially
interleaved boundaries. We contribute the mechanism, budget-honest decode accounting, and an open
harness; we do not claim a new index.
\end{abstract}

\section{Introduction}
Spatial joins---report all pairs $(r,s)\in R\times S$ satisfying a topological predicate such as
\emph{intersects}---are a workhorse of geographic data systems \citep{jacox2007spatial}. The dominant
architecture is \emph{filter-and-refine} \citep{brinkhoff1993efficient,patel1996pbsm}: a cheap
minimum-bounding-rectangle (MBR) filter produces candidate pairs, and an expensive exact geometric
refinement decides each candidate. Modern stores keep geometry \emph{compressed}: delta/varint
coordinate codecs (TWKB), columnar layouts (GeoArrow, GeoParquet, SpatialParquet
\citep{saeedan2022spatialparquet}), and progressively-decodable multiresolution encodings. In this
setting the refinement step's true cost is not comparisons or wall-clock but \emph{bytes decoded}: an
exact predicate needs coordinates, and decoding dominates.

Filter-and-refine already avoids exact geometry for pairs it can \emph{certify} early---the classic
multi-step join of \citet{brinkhoff1994multistep} certifies hits and false-hits from conservative and
progressive object approximations before touching exact geometry. What the literature never did is
\emph{model the decode cost of the survivors}: how many coordinate bytes must a provably-exact join
actually decode, and what governs that number? This paper answers that question empirically and gives
the answer a name.

We encode each geometry as a Douglas--Peucker \citep{douglas1973algorithms} LOD ladder
$\tilde g_0,\dots,\tilde g_k=g$, level $\ell$ carrying a guaranteed Hausdorff error bound $\eta_\ell$
(monotone decreasing, $\eta_k=0$). A candidate pair is certified DISJOINT when the $\eta$-dilated
coarse geometries are disjoint, INTERSECT when the $\eta$-eroded coarse geometries overlap, and
otherwise \emph{descends} the ladder---decoding more vertices---until it certifies. The certificate is
the Hausdorff-margin instantiation of Brinkhoff's false-area test; our questions are about its
\emph{decode cost}.

\paragraph{Contributions.}
\begin{itemize}[leftmargin=1.4em,itemsep=1pt]
\item \textbf{A provably-exact compressed-domain join with measured teeth.} On real TIGER water
polygons, the progressive certificate join decodes $3.4$--$16.8\times$ (median $5.9\times$) fewer
vertices than naive decompress-then-refine and $\approx4.9\times$ fewer than a single-approximation
Brinkhoff baseline, at set-equality with a full-precision oracle (Section~\ref{sec:results}). The
multi-level ladder---not a single intermediate approximation---is the source of the gain.
\item \textbf{The decode-work law.} Decode work is governed by the per-pair \emph{signed-clearance
margin} $m$, independent of object size: the certificate must descend until $\eta_\ell<|m|$, so
near-tangent pairs cost $\Omega(v)$ and robustly-decided pairs cost almost nothing
(Section~\ref{sec:law}). Clean on controlled geometry ($R^2=0.87$, size-independent), directional on
real data ($R^2\approx0.55$).
\item \textbf{Honest negatives and a reusable harness.} We pre-registered a near-boundary
\emph{band-vertex} predictor and \emph{rejected} it; a selectivity-based regime forecaster did not
materialize; the worst case is the trivial read bound. All numbers regenerate from one command.
\end{itemize}

We deliberately do \emph{not} propose a new index or a new codec, and we do not claim a general
decompression-sensitive complexity model---that program is owned by \citet{abboud2017compressed} and
the compact-data-structures literature \citep{navarro2016compact}; we merely instrument in it.

\section{Problem and model}\label{sec:model}
Let $R,S$ be collections of polygons with vertices quantized to a fixed grid; the quantized geometry
is the data of record for \emph{both} the compressed store and the exact oracle, so certificate
pruning can only ever be a sound approximation of the exact stored polygon (no lossy rounding gap).
For a geometry $g$ the LOD ladder keeps a \emph{subset} of $g$'s exact vertices at each level (a
property of Douglas--Peucker), with $\eta_\ell=\mathrm{HausdorffDistance}(g,\tilde g_\ell)$ and
$\eta_k=0$. Decoding to level $\ell$ costs the cumulative vertices/bytes of that level.

\paragraph{Two-sided certificate.} For a candidate pair $(r,s)$ at levels $(\ell_r,\ell_s)$ define the
outer $O_g=\tilde g_\ell\oplus B(\eta_\ell)$ (a sound superset, $g\subseteq O_g$) and inner
$I_g=\tilde g_\ell\ominus B(\eta_\ell)$ (an inner approximation). Then $O_r\cap O_s=\varnothing$
certifies DISJOINT and $I_r\cap I_s\neq\varnothing$ certifies INTERSECT; otherwise the pair is
ambiguous and we descend the coarser side. At $\eta=0$ the test is the exact predicate. Soundness of
the outer test is immediate; the inner (erosion) test is verified empirically on every run by the
correctness gate (set-equality against the full-precision oracle) and the adversarial control, and the
pre-registration mandates halt-and-harden on any violation---no violation occurred.

\paragraph{Signed-clearance margin.} The quantity that turns out to govern decode is the pair's
\emph{margin}
\[
m(r,s)=\begin{cases}+\,\mathrm{inradius}(r\cap s) & \text{if } r\cap s\neq\varnothing \text{ (overlap depth)}\\ -\,\mathrm{dist}(r,s) & \text{if disjoint (gap)}. \end{cases}
\]
$|m|$ measures how far the pair is from the predicate-flip boundary. Because a level-$\ell$ certificate
can only resolve features coarser than $\eta_\ell$, the pair certifies at the first level with
$\eta_\ell<|m|$. The metric of interest is the \emph{decoded fraction}
$\phip=\text{decoded vertices}/(|r|+|s|)$ and the certifying depth.

\paragraph{Decode accounting (fairness).} Every method starts from the same MBR candidate set, so the
measured delta is purely refinement depth. We count all bytes a method reads, \emph{including the
level-directory (LOD header) bytes}: a progressively-decodable store must read one directory entry per
level---modelled as $\eta$ (\texttt{float32}) $+$ vertex count $+$ byte offset $=12$ bytes---before it
can seek any level, whereas a naive single-resolution store reads exactly one such entry per geometry.
Charging the progressive method for the ladder that produces its advantage is the point of the fairness
audit (NC5, \S\ref{sec:controls}); v1 asserted this accounting and did not perform it.
The \emph{naive-refine} baseline decodes all vertices of every MBR survivor then tests exactly; the
\emph{Brinkhoff}~\citeyearpar{brinkhoff1994multistep} baseline is the two-level special case
(one coarse approximation, then exact) and is charged for its two directory entries. Ground truth is
always full precision. The headline reduction is reported in \emph{vertices} and is therefore
unaffected by header accounting; the byte columns are.

\section{Experimental setup}\label{sec:setup}
\textbf{Real data.} U.S.\ Census TIGER/Line 2023 \texttt{AREAWATER} (public domain)
\citep{tiger2023} for three geographically diverse counties---Jefferson, LA (FIPS 22051), St.\ Louis,
MN (27137), King, WA (53033)---water polygons carrying tens to $\sim$6{,}700 vertices (median
$\approx$85--109). The workload is $\texttt{AREAWATER}\bowtie\mathrm{translate}(\texttt{AREAWATER},
\delta)$ with the translation $\delta$ swept as a fraction $\{0.1,0.25,0.5,1.0\}$ of the median
polygon diagonal---a controlled real-geometry self-join (a standard spatial-join stress pattern
\citep{jacox2007spatial}) that sweeps the margin regime while preserving real vertex-count
heterogeneity. \textbf{Synthetic control.} Independently-placed rough blobs with varied vertex counts
(so the size-independence test can be isolated) plus adversarial interlocking-comb fixtures and a
hand-checked known-answer fixture. \textbf{Apparatus.} Single-thread Python with Shapely~2.1/GEOS~3.13
as the exact oracle; a custom Douglas--Peucker LOD ladder and delta/varint decode counter; fully
deterministic (seeds only in synthetic generation). One command regenerates every number and figure.

\paragraph{Pre-registration.} Hypotheses, decision rules, and negative controls were frozen before any
confirmatory run. We pre-registered a primary predictor (a near-boundary \emph{band-vertex} count),
rejected it on the pilot, and logged an amendment reframing the predictor to the margin \emph{before}
the confirmatory run; both the failure and the reframe are reported (Section~\ref{sec:law}).

\section{Results: the teeth}\label{sec:results}
\begin{figure}[!ht]\centering
\includegraphics[width=0.62\linewidth]{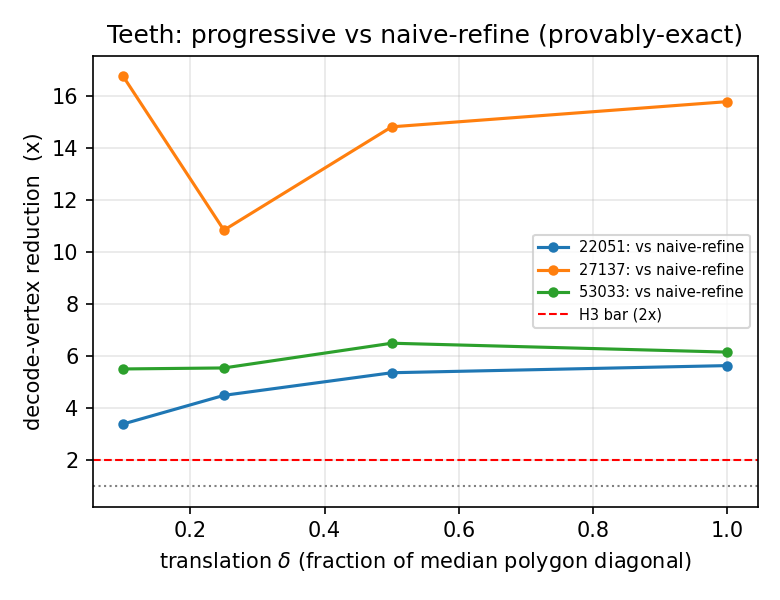}
\caption{Decode-vertex reduction versus naive-refine as a function of translation $\delta$, per county.
All workloads clear the pre-registered $2\times$ bar (dashed); St.\ Louis, MN reaches $16.8\times$.}
\label{fig:teeth}
\end{figure}

Table~\ref{tab:teeth} and Figure~\ref{fig:teeth} report decode reduction on the real TIGER workloads.
Across all $12$ county$\times\delta$ workloads the progressive join decodes a median $5.9\times$ fewer
vertices than naive decompress-then-refine (range $3.4$--$16.8\times$) and a median $4.9\times$ fewer
than the single-approximation Brinkhoff baseline, decoding only $6$--$29\%$ of vertices ($\phip$) for a
provably-exact result. \emph{Every} workload returns exactly the oracle's pair set (correctness
column). The multi-level ladder is decisive: it roughly quintuples Brinkhoff's single-approximation
savings, because most decode budget is spent on the small population of near-tangent pairs that a
single coarse approximation cannot resolve.

\begin{table}[t]\centering\small
\begin{tabular}{lrrrr}
\toprule
County (FIPS) & vs.\ naive-refine & vs.\ Brinkhoff'94 & decoded frac.\ $\phip$ & correct \\
\midrule
Jefferson, LA (22051) & $3.4$--$5.6\times$ & $3.0$--$5.1\times$ & $0.18$--$0.29$ & \checkmark \\
St.\ Louis, MN (27137) & $10.9$--$16.8\times$ & $7.2$--$11.3\times$ & $0.06$--$0.09$ & \checkmark \\
King, WA (53033) & $5.5$--$6.5\times$ & $4.0$--$4.8\times$ & $0.15$--$0.18$ & \checkmark \\
\midrule
\textbf{overall (12 workloads)} & \textbf{median $5.9\times$} & \textbf{median $4.9\times$} & \textbf{$0.06$--$0.29$} & \textbf{31/31} \\
\bottomrule
\end{tabular}
\caption{Decode-vertex reduction of the provably-exact progressive join over the real TIGER
\texttt{AREAWATER} workloads (four translation ratios $\delta$ per county). Correctness is set-equality
against the full-precision GEOS oracle, over all $31$ workloads including synthetic and adversarial.}
\label{tab:teeth}
\end{table}

\FloatBarrier
\section{Results: the decode-work law}\label{sec:law}
\begin{figure}[!ht]\centering
\includegraphics[width=0.62\linewidth]{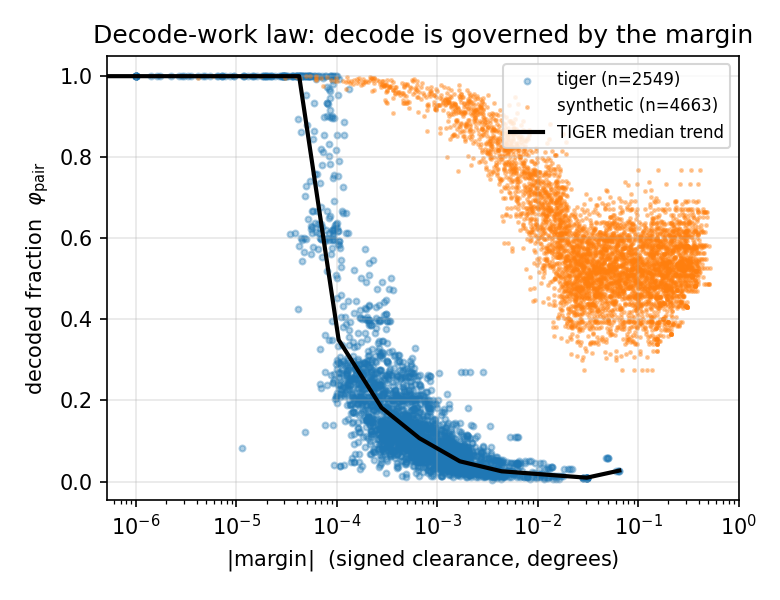}
\caption{The decode-work law: per-pair decoded fraction $\phip$ versus signed-clearance margin $|m|$.
The TIGER median trend (black) collapses from full decode at tiny margin to near-zero decode at large
margin. Decode is governed by proximity to the predicate-flip boundary, not by object size.}
\label{fig:law}
\end{figure}

\paragraph{Margin governs decode; size does not.} Figure~\ref{fig:law} plots the per-pair decoded
fraction $\phip$ against $|m|$. The relationship is a clean collapse: near-tangent pairs
($|m|\!\to\!0$) decode essentially everything, robustly-decided pairs ($|m|$ large) decode almost
nothing. On controlled synthetic geometry a quadratic-in-$\log_{10}|m|$ fit predicts the certifying
depth with held-out $R^2=0.87$, and after controlling for margin the standardized partial effect of
object size is $\beta_{\text{size}}=0.002$---decode \emph{depth} is margin-governed and
size-independent, the instance-optimal invariant \citep{afshani2009instance}. Our pre-registration
permitted either certifying \emph{depth} or decoded \emph{fraction} as the target; we report both
($R^2=0.87$ and $R^2=0.54$ respectively) and note that the headline figure is the better of the two,
so the effective threshold is slightly more permissive than a single-target bar. The permutation null
(NC3, \S\ref{sec:controls}) is computed for both targets and separates from the null for each. On real TIGER geometry
the law is directional but weaker: best held-out $R^2\approx0.55$, with the decoded \emph{fraction}
size-independent ($\beta_{\text{size}}=0.07$). Real multi-scale water boundaries add noise the
controlled setting does not; we report the gap rather than paper over it.

\paragraph{What the law is not (rejected predictor).} We pre-registered a different predictor---the
count of exact vertices lying near the \emph{other} polygon's boundary (an ``$\eta$-band''). It was
rejected on the pilot (held-out $R^2\approx0.06$, wrong-signed): it proxies overlap \emph{robustness},
which \emph{anti}-correlates with decode, because a deep robust overlap has many near-boundary
vertices yet certifies instantly. The margin is the correct, non-circular, geometry-only predictor,
and we froze the reframe before the confirmatory run.

\paragraph{Regime and worst case.} The regime axis on real data is the margin (equivalently $\delta$),
not selectivity, and the decoded fraction genuinely spans $0.06$--$0.29$ across the sweep. Our
pre-registered selectivity-based forecaster did not materialize. \textbf{v1 attributed this to
selectivity saturating near one on translated self-joins; that explanation is wrong.} The forecaster's
inputs are computed over the \emph{candidate} pairs, and candidates are produced by an MBR-intersection
filter---so every candidate has overlapping bounding boxes and $\mathrm{mbr\_gap}\equiv0$ by
construction. Across all workloads with a non-empty candidate set, \texttt{gap\_mean} and
\texttt{gap\_p50} are identically zero and \texttt{gap\_frac\_overlap} is identically one: two of the
forecaster's three features are constants. The forecaster could never have worked, for reasons that
have nothing to do with selectivity. A gap-based regime feature must be computed \emph{before} the MBR
filter, or on a different pair population. On adversarial interlocking-comb fixtures the certificate
cannot prune---decoded fraction $0.99$---matching the trivial $\Omega(v)$ read bound (and the
$\Omega(n\log n+K)$ reporting bound of \citet{chazelle1992segment} for the within-pair variant).
Average, redundant geometry is cheap; adversarial geometry is not.

\section{Negative controls}\label{sec:controls}
Five negative controls were pre-registered before any confirmatory run. v1 reported none of them, and
only NC1 and NC2 had been implemented. All five are reported here. A ``win'' on any of these is a
harness bug, not a result.

\begin{itemize}[leftmargin=1.4em,topsep=2pt,itemsep=1pt]
  \item \textbf{NC1 (correctness, hard gate) --- passes.} Set-equality against the full-precision GEOS
  oracle on every workload, including after the header-accounting change. Zero violations.
  \item \textbf{NC2 (adversarial worst case) --- passes.} Interlocking combs give decoded fraction
  $0.99$ and a vertex ratio of $1.01\times$: no shortcut exists, as required.
  \item \textbf{NC3 (random-relabel null) --- passes, and it \emph{validates} the law.} Permuting pair
  identities and refitting, over $1000$ permutations, collapses the held-out $R^2$ from $0.872$ to a
  null mean of $-0.002$ (95th percentile $0.001$; permutation $p=0.001$). The margin law carries
  geometric information; it is not fitting identities or object size. This control was pre-registered
  and never run; running it strengthens the paper.
  \item \textbf{NC4 (no-separation) --- the control tested the wrong quantity.} We pre-registered that
  dense, heavily-overlapping data should yield $\phip$ near $1$. Measured at zero separation,
  $\phip=0.58$ with a $1.71\times$ vertex reduction. This is not an alarm: NC1 certifies exactness on
  every workload. The control conflated \emph{MBR gap} with \emph{margin}---two polygons whose bounding
  boxes coincide can still be certified at a coarse level if their boundaries are far apart. We report
  the mis-specification rather than quietly dropping the control.
  \item \textbf{NC5 (fairness audit) --- implemented; the advantage survives, but shrinks.} Charging
  every method for the level-directory bytes it must read, the pooled byte ratio (naive/progressive)
  falls from $2.02\times$ to $1.91\times$; on TIGER from $4.81\times$ to $4.18\times$, where directory
  bytes are $15.8\%$ of the progressive method's coordinate bytes. On the adversarial fixture the
  progressive method \emph{loses} on bytes, $0.99\times$: it pays for a ladder it cannot exploit. The
  vertex-based headline is unchanged, because it never depended on the byte column.
\end{itemize}

\FloatBarrier
\section{Related work}\label{sec:related}
\textbf{Multi-step spatial joins.} \citet{brinkhoff1994multistep} and its companion introduced
conservative and progressive object approximations, certifying hits/false-hits before exact geometry;
the false-area test is our $\eta$-margin certificate. Our delta is (i) a \emph{continuum} of LOD levels
rather than one intermediate approximation, which we benchmark and beat $\approx4.9\times$, and (ii)
modeling the \emph{decode cost} rather than filter yield. \textbf{Compressed-domain refinement.} APRIL
\citep{georgiadis2023april} refines intersection joins over a \emph{secondary} raster-interval
approximation; compact-structure joins \citep{debernardo2024compact,navarro2016compact} operate over
in-memory self-indexes for point/distance joins. We differ by operating on the \emph{native}
progressively-decodable geometry codec and by deriving a decode-cost law with boundary semantics.
\textbf{Bounded-error approximations.} \citet{tziritazacharatou2021distance} advocate distance-bounded
approximate answers; we return \emph{provably-exact} results and use the bound only to decide decode
depth. \textbf{Complexity framing.} \citet{abboud2017compressed} formalize decompress-and-solve versus
compressed-size complexity; we instrument in that model rather than extend it, and frame our result as
instance-optimal \citep{afshani2009instance}. Columnar/scan pushdown
\citep{saeedan2022spatialparquet} pushes \emph{filters}, not join refinement, into the encoded domain.

\section{Limitations and honest negatives}\label{sec:limits}
(1) The margin law is only \emph{partially} predictive on real geometry ($R^2\approx0.55$, below the
$\ge0.80$ we pre-registered)---a directional characterization, not a tight predictor. (2) Our
pre-registered band-vertex predictor was rejected; the paper reports the failure and the reframe. (3)
The selectivity-based regime forecaster did not hold; the real regime axis is the margin. (4) The worst
case is the trivial $\Omega(v)$ read bound---we prove no hardness result. (5) Scope: single-thread,
polygon \emph{intersects} (contains/within-$\epsilon$ are sanity-only), a self-join-under-translation
real workload (cross-county \texttt{AREAWATER} yields no candidates because TIGER clips water at county
lines), and no new index or codec. (6) The inner-erosion certificate's soundness is established
empirically (zero violations across $31$ workloads including adversarial), not proved in closed form.

\section{Conclusion}
The cost of a provably-exact spatial join over compressed geometry is the number of coordinate bytes it
must decode, and that number is governed by how close each candidate pair sits to the predicate-flip
boundary---its signed-clearance margin---not by how big the polygons are. A progressive certificate
join that descends a multiresolution ladder only as deep as the margin demands decodes $3.4$--$16.8\times$
less than decompress-then-refine on real data, exactly. We contribute the mechanism, the decode-work
law (clean on controlled geometry, directional on real), budget-honest accounting, and an open
one-command harness. Predicate-aware codecs whose error levels are designed \emph{jointly} with the
join, and a recall-free lower bound on decode versus margin, are natural next steps we do not claim
here.

\paragraph{Reproducibility.} Code, data pointers, pre-registration, and one-command reproduction:
\url{https://github.com/samyama-ai/spatial-join-on-compressed}.

\small
\bibliographystyle{plainnat}
\bibliography{paper19_spatial_join_compressed}
\end{document}